\documentclass[prb,twocolumn,showpacs,preprintnumbers,amsmath,amssymb]{revtex4}

\usepackage{graphicx}
\usepackage{bm}
\usepackage{color}
\usepackage{amsmath}
\usepackage{amssymb}

\begin{document}


\title{Effects of dilution on magnetic and transport properties of
 La$_{0.7}$Ca$_{0.3}$Mn$_{1-x}$\textit{M}$^\prime_x$O$_3$}
\author {D. N. H. Nam}
\email{daonhnam@yahoo.com}
\affiliation{Institute of Materials Science, VAST, 18 Hoang-Quoc-Viet, Hanoi, Vietnam}
\author {H. S. Hong}
\author {N. V. Khien}
\affiliation{College of Technology, Hanoi National University, Hanoi, Vietnam}
\affiliation{Institute of Materials Science, VAST, 18 Hoang-Quoc-Viet, Hanoi, Vietnam}
\author {N. V. Dai}
\author {T. D. Thanh}
\author {L. T. C. Tuong}
\author {L. V. Hong}
\author {N. X. Phuc}
\affiliation{Institute of Materials Science, VAST, 18 Hoang-Quoc-Viet, Hanoi, Vietnam}
\date{\today}
%
\begin{abstract}
Magnetic and transport properties of La$_{0.7}$Ca$_{0.3}$Mn$_{1-x}$\textit{M}$^\prime_x$O$_3$ (\textit{M}$^\prime$=Al,Ti) are studied. The dilution of the Mn lattice results in a weakening of the ferromagnetism, a deterioration of the metallic conductivity, and a strong enhancement of the magnetoresistance. Although $T_c$ linearly decreases with $x$ in the low substitution ranges for both \textit{M}$^\prime$ series, the scaling behavior $T_c(n_p)$ previously observed for La$_{0.7}$Sr$_{0.3}$Mn$_{1-x}$\textit{M}$^\prime_x$O$_3$ [Phys. Rev. B 73, 184403 (2006)] is no longer obtained. Extrapolations of the $T_c(n_p)$ linear curves to $T_c$=0 give $n_p$ values much smaller than 1. These results suggest that, according to a molecular-field approximation, antiferromagnetic superexchange between Mn ions is significant in La$_{0.7}$Ca$_{0.3}$MnO$_3$, in contrast to what observed in La$_{0.7}$Sr$_{0.3}$MnO$_3$. Additionally, structural data of the Al-substituted samples suggest that variations of the $e_g$-electron bandwidth $W$ cannot explain the decrease of $T_c$ in magnetically diluted manganites.
\end{abstract}
\pacs{75.10.Hk, 75.30.Et, 75.47.Gk, 75.47.Lx}
\maketitle

\section{Introduction}
Mixed-valance manganites $(R,A)$MnO$_3$ ($R$: rare-earth, $A$: alkaline-earth elements) have been intensively studied for more than a decade since the discovery of the Colossal MagnetoResistance (CMR) phenomenon.\cite{Kusters,Helmolt,Jin} This class of materials is attractive to scientists due to not only their potential for practical applications, but also their rich and intriguing fundamental physics. The Mn$^{3+}$/Mn$^{4+}$ mixed valence leads to a coexistence of competing interactions among the Mn ions. While antiferromagnetic (AF) and insulator-like behavior are found in the systems dominated by Mn$^{3+}$-O$^{2-}$-Mn$^{3+}$ and Mn$^{4+}$-O$^{2-}$-Mn$^{4+}$ superexchange (SE) interactions, ferromagnetic (FM) and metallic behavior are realized in those Mn$^{3+}$-O$^{2-}$-Mn$^{4+}$ double-exchange (DE) couplings\cite{Zener} are pertinently introduced by doping $A^{2+}$ ions at the $R$-site. Along with the mixed interaction character, lattice distortions (such as Jahn-Teller and GdFeO$_3$-type) and phase segregation phenomena also play important roles governing the physical properties of the materials. Depending on the compositions and external conditions, the properties of manganites can be ranged from disordered to ordered, from antiferromagnetic to ferromagnetic, from insulating to metallic.\cite{Coey,Dagotto}

On the search for compositions with novel properties, results for both $R$- and Mn-site substitutions have been quite well documented in the literature. For the $R$-site substitution, magnetic and transport properties have been found to be quite systematically influenced by the GdFeO$_3$-type lattice distortion and lattice disorder associated with the size and size mismatch of the $R$-site cations.\cite{Hwang,Teresa,Arulraj} However, doping at the Mn-site would create a much more complicated scenario. Since double exchange has been found to exist so far only between Mn ions, any substitution at the Mn-site would lead to a weakening of the DE ferromagnetism and therefore a deterioration of the metal-like conduction. If the substituting ion carries a net spin moment, magnetic interactions between Mn ions and the substituent are inevitable. The exchange interactions among Mn ions would also vary with doping as a result of the structural modification due to ionic size mismatch. In addition, the ionic ratio Mn$^{3+}$/Mn$^{4+}$ would be changed adapting to the valence and concentration of the substituting ions. These effects therefore effectively hinder systematic studies of the Mn-site substitution in manganites.

In a recent study,\cite{Nam} it has been found that a selective substitution of Mn by non-magnetic elements could reveal important information on the interactions among Mn ions in a manganite. The authors proposed that the dilution of the Mn lattice leads to a decrease of the molecular field acting on Mn ions and therefore suppresses the ferromagnetism. In the low substitution ranges, La$_{0.7}$Sr$_{0.3}$Mn$_{1-x}$\textit{M}$^\prime_x$O$_3$ (\textit{M}$^\prime$=Al,Ti) exhibits a linear scaling behavior between $T_c$ and the dilution concentration $n_p$ of \textit{M}$^\prime$ ions ($n_p$=$x$/0.7 for \textit{M}$^\prime$=Al and $x$/0.3 for \textit{M}$^\prime$=Ti). Since $T_c$ has the tendency to reduce to zero at $n_p$=1, based on the molecular-field theory (MFT) approximation, the authors suggested that the DE mechanism is totally dominant in La$_{0.7}$Sr$_{0.3}$MnO$_3$, playing a role behind the fact that the system has the highest $T_c$ among perovskite manganites known to date. In the present paper, we report our study on the magnetic and transport properties of La$_{0.7}$Ca$_{0.3}$Mn$_{1-x}$\textit{M}$^\prime_x$O$_3$ (\textit{M}$^\prime$=Al,Ti) compounds. Results from previous studies on similar systems were reported and can be referenced in a number of publications.\cite{Turilli,Blasco,Li,LiuX,LiuYH,Cao,Lakshmi} Our study is focused mainly on the dilution effects in La$_{0.7}$Ca$_{0.3}$Mn$_{1-x}$\textit{M}$^\prime_x$O$_3$ in comparison with La$_{0.7}$Sr$_{0.3}$Mn$_{1-x}$\textit{M}$^\prime_x$O$_3$. The dilution of Mn lattice results in a strong suppression of the ferromagnetism, a deterioration of the conductivity, and a huge development of the magnetoresistance. The $T_c(n_p)$ curves obtained for the two series of our samples, according to the MFT approximation, suggest a significant contribution of antiferromagnetic superexchange in La$_{0.7}$Ca$_{0.3}$MnO$_3$, in contrast to La$_{0.7}$Sr$_{0.3}$MnO$_3$.
\section{Experiment}
All the La$_{0.7}$Ca$_{0.3}$Mn$_{1-x}$\textit{M}$^\prime_x$O$_3$ (denoted as LCMA$_x$ for \textit{M}$^\prime$=Al and LCMT$_x$ for \textit{M}$^\prime$=Ti) samples were prepared using a conventional solid state reaction method from pure ($\geqslant$99.99\%) raw powders of La$_2$O$_3$, CaCO$_3$, MnO$_2$, Al$_2$O$_3$, and TiO$_2$. The powders with appropriate amounts were thoroughly ground, mixed, pelletized, and then calcined at several processing steps with
increasing temperatures from 900 $^\mathrm{o}$C to 1200 $^\mathrm{o}$C and intermediate grindings. The products were then sintered at 1300 $^\mathrm{o}$C for 48 h in ambient atmosphere followed by a very slow cooling process from the sintering to room temperature with an annealing step at 700 $^\mathrm{o}$C for 5 hours. Room-temperature x-ray diffraction patterns (measured by a SIEMENS-D5000 with Cu-$K_\alpha$ radiation) showed that all of the samples are single phase with perovskite orthorhombic (space group $pnma$) structures; structural data were calculated using Rietveld refinements. Redox titration experiments (using K$_2$Cr$_2$O$_7$ titrant and C$_{24}$H$_{20}$BaN$_2$O$_6$S$_2$ colorimetric indicator) show almost no oxygen deficiencies or excesses ($\delta$$\leq$0.006) in all the samples. Magnetic and four-probe resistance/magnetoresistance measurements were carried out in a Quantum Design PPMS6000 system.
\section{Results and Discussion}
\begin{figure}[b!]
  \includegraphics[width=6.3cm]{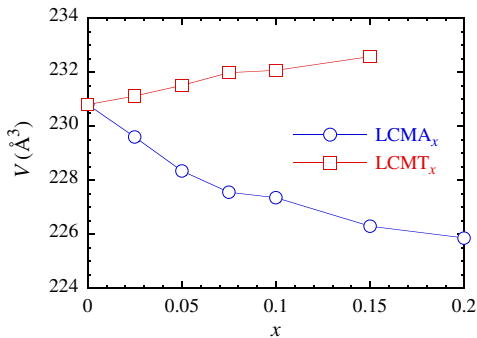}
  \caption{(Color online) Variations of the unit cell volume with Al and Ti concentrations. This result demonstrates an additional evidence for the selective substitutions of Al$^{3+}$ and Ti$^{4+}$ for Mn$^{3+}$ and Mn$^{4+}$, respectively.}\label{Fig1}
\end{figure}
In several previous publications, there were concerns that Al$^{3+}$ could substitute for Mn$^{4+}$, resulting in an oxygen understoichiometry in some Al-doped manganites.\cite{Blasco,Krishnan,Nair} In the case of Ti-doping, an oxygen overstoichiometry might be realized if Mn$^{3+}$ is replaced with Ti$^{4+}$. If the oxygen concentration is kept stoichiometric, the substitution of Al and Ti for Mn is therefore expected to be selective, implying that Al$^{3+}$ would only substitute for Mn$^{3+}$ and Ti$^{4+}$ for Mn$^{4+}$.\cite{Nam,LiuX} This is indeed supported in the present La$_{0.7}$Ca$_{0.3}$Mn$_{1-x}$\textit{M}$^\prime_x$O$_3$ samples by the fact that there is almost no deficient or excessive oxygen as determined from the redox titration experiments. Moreover, the plots in Fig. \ref{Fig1} show that the unit cell volume derived from x-ray diffraction data monotonically decreases with Al substitution while it increases with increasing Ti concentration. These structural variations give a further convincing evidence for the selective substitution of Al$^{3+}$ and Ti$^{4+}$ for Mn$^{3+}$ and Mn$^{4+}$, respectively, considering that the ionic radius of Al$^{3+}$ (0.535 {\AA}) is smaller than that of Mn$^{3+}$ (0.645 {\AA}) and Ti$^{4+}$ (0.605 {\AA}) is larger than Mn$^{4+}$ (0.530 {\AA}).\cite{Shannon}
\begin{figure}
  \includegraphics[width=6.3cm]{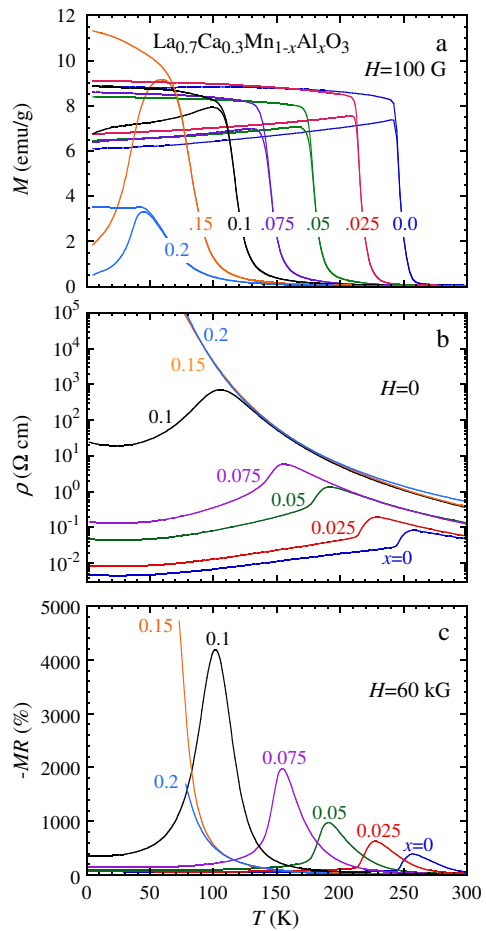}
  \caption{(Color online) LCMA$_x$: (a) $M_\mathrm{ZFC}(T)$ and $M_\mathrm{FC}(T)$ curves measured in $H$=100 G, (b) $\rho(T)$ curves in zero field, and (c) \textit{-MR}($T$) curves in $H$=60 kG.}\label{Fig2}
\end{figure}

Measurements of temperature dependent magnetization, $M(T)$, are carried out for all the samples in both zero-field-cooling (ZFC) and field-cooling (FC) protocols. Fig. \ref{Fig2}a presents the $M(T)$ curves of the LCMA$_x$ series. A sharp ferromagnetic-paramagnetic (FM-PM) phase transition is observed for all the samples in the low doping region. The parent compound, La$_{0.7}$Ca$_{0.3}$MnO$_3$, shows a FM-PM transition temperature $T_c$ of $\sim$247 K. With increasing $x$, the ferromagnetism is severely suppressed, indicated by a drastic decrease of $T_c$. The shape of the $M_\mathrm{ZFC}(T)$ and $M_\mathrm{FC}(T)$ curves of the $x$$\geqslant$0.15 samples suggests that, although a FM-PM transition is still observed for $x$=0.15, they are no longer pure ferromagnets, but spin-glass-like systems of short-range ferromagnetic correlations. The $T_c$ vs $x$ data extracted from the $M(T)$ curves for the ferromagnetic samples are plotted in the inset of Fig. \ref{Fig5} and will be discussed later in detail.

As can be seen in Fig. \ref{Fig2}b, the conductivity is also strongly suppressed by the Al substitution. Although a metal-insulator (MI) transition is observed for all the ferromagnetic ($x$$\leqslant$0.1) compounds near the transition temperature $T_c$, the resistivity increases in both FM and PM states. The resistivity drop at $T_c$ and the metallic conducting behavior in the FM state can be well explained by the double-exchange (DE) mechanism where a parallel alignment of the magnetic moments favors the transfer of $e_g$ electrons between Mn$^{3+}$ and Mn$^{4+}$ ions. Since Al$^{3+}$ does not participate in double exchange, it blocks all the conducting paths at the sites it occupies. The metallic conducting behavior is thus not expected in the highly substituted samples where the long-range FM order collapses into short-range FM correlations; the system consists of conducting FM regions separated by insulating non-FM boundaries. Such behavior can be seen in the $x$=0.15 and 0.2 samples as the resistance with decreasing temperature monotonically increases up to the limit of our measurement system ($\sim$2$\times10^6$ $\Omega$) showing no observable MI transition. The resistance of these samples is still unmeasurably high down to temperatures as low as of 2 K and even in an applied field of 60 kG. The magnetoresistance ratios \textit{MR}, defined as \textit{MR}(\%)=($R_H$-$R_0$)$\times$100/$R_H$, measured in an applied field $H$=60 kG for all the LCMA$_x$ compounds are presented in Fig. \ref{Fig2}c. As expected, all the low doped samples show a peak of \textit{MR} near the FM-PM phase transition that is typical of the CMR effect. Although a maximum of \textit{MR} cannot be measured for the highly doped samples, $x$=0.15 and 0.2, a clear development of the CMR effect with increasing Al concentration is observed. At the phase transition, the \textit{MR} value increases from $\sim$400\% for $x$=0 to $\sim$4200\% for $x$=0.1. The $x$=0.15 sample shows a monotonous increase of \textit{MR} with lowering temperature that reaches up to $\sim$4700\% at $\sim$73 K. A similar improvement of the CMR effect was previously observed in La$_{0.67}$Sr$_{0.33}$Mn$_{1-x}$Al$_{x}$O$_3$ by Turilli \textit{et al.}.\cite{Turilli} The authors proposed a phenomenological model to quantitatively explain the relationship between the magnetic and transport characteristics. Later, Blasco \textit{et al.}\cite{Blasco} reported a huge development of magnetoresistance (\textit{MR}=$10^7$\% in $H$=12 T) of La$_{2/3}$Ca$_{1/3}$Mn$_{1-x}$Al$_x$O$_{3\pm\delta}$. Since their Al-doped samples are not homogeneous, the authors attribute the increase of \textit{MR} to the effect of disorder in the insulating regions. In our work, all the ferromagnetic samples ($x$$\geqslant$0.1) exhibit a MI transition and a maximum of \textit{MR} at temperatures very close to the FM-PM phase transitions. Thus, the increase of \textit{MR} should not be attributed to magnetic field effect on the insulating regions, but directly to the FM ordering according to the DE mechanism. The partial substitution of Al for Mn not only locally blocks conducting paths where Al occupies, but also globally weaken the FM order of the Mn lattice; both lead to an increase of $R_0$. The later case is directly associated with the decrease of $T_c$ and the enhancement of \textit{MR} at $T_c$ in the low doped samples. For highly doped ones ($x$$\geqslant$0.15), the increase of \textit{MR} with lowering temperature is due to an expansion of ferromagnetic regions under high magnetic field. If the Al concentration is high enough to prevent the ferromagnetic regions from percolating throughout the sample, the magnetoresistance would increase monotonically with lowering temperature. However, too high Al concentration would also suppress the growth of FM regions and therefore reduce the magnetoresistance effect.
\begin{figure}
  \includegraphics[width=6.3cm]{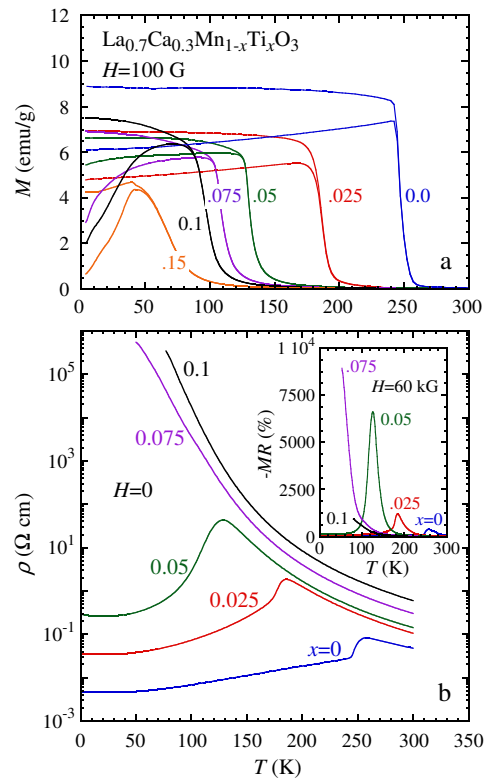}
  \caption{(Color online)  LCMT$_x$: (a) $M_\mathrm{ZFC}(T)$ and $M_\mathrm{FC}(T)$ curves measured in $H$=100 G, (b) $\rho(T)$ curves in zero field, and the inset of (b) shows \textit{-MR}($T$) curves in $H$=60 kG.}\label{Fig3}
\end{figure}

Figure \ref{Fig3} presents the magnetization and transport/magnetotransport characterizations of the LCMT$_x$ compounds. Qualitatively, the Ti substitution causes similar effects on the magnetic and transport properties as those caused by the Al substitution. The substitution suppresses $T_c$, increases resistivity in both PM and FM states, and strongly enhances the magnetoresistance effect. A sharp FM-PM phase transition with a well-defined $T_c$ is observed for all the samples at low substitution concentrations. Although a PM-FM phase transition is still observed for $x$=0.075, the insulating behavior (Fig. \ref{Fig3}b) suggests that this compound is no longer a pure ferromagnet, but a system consisting of conducting FM regions separated by insulating non-FM boundaries. The highest substituted sample, $x$=0.1, shows the behavior of a spin-glass-like insulating system. While all the FM samples, $x$$\leqslant$0.05, that exhibit a MI transition concomitant with a maximum of \textit{MR} at $T_c$, those with insulating behavior exhibit a monotonous increase of \textit{MR} with lowering temperature (see the inset of Fig. \ref{Fig3}b). The \textit{MR} value for $x$=0.075 reaches $\sim$9000\% at 50 K in an external field of 60 kG.
\begin{figure}
  \includegraphics[width=6.3cm]{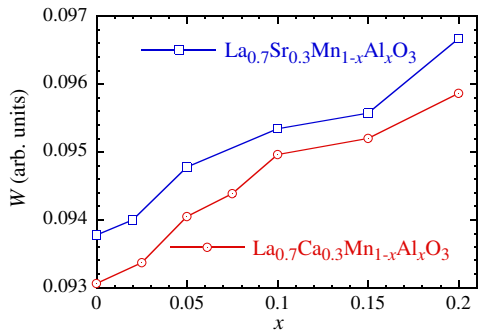}
  \caption{(Color online) Variations of the conduction electron bandwidth $W$ with Al and Ti concentrations. The monotonic increase in $W(x)$ is contradictory to the reduction in $T_c(x)$ in both La$_{0.7}$Ca$_{0.3}$Mn$_{1-x}$Al$_x$O$_3$ and La$_{0.7}$Sr$_{0.3}$Mn$_{1-x}$Al$_x$O$_3$ systems.}\label{Fig4}
\end{figure}

 There have been several explanations for the reduction of $T_c$ in ferromagnetic manganites where Mn is substituted by a trivalent or quadrivalent non-magnetic element such as Al and Ti. One explanation is simply that this is due to the suppression of long-range FM order of the localized $t_\mathrm{2g}$ spins by local breakdown of the exchange couplings where the substitution occurs.\cite{Qin,Sawaki} Hu \textit{et al.} \cite{Hu} assumed a demolition of the DE Mn$^{3+}$-O$^{2-}$-Mn$^{4+}$ bonds and a lower hole-carrier concentration caused by Ti substitution. Kallel \textit{et al.} \cite{Kallel} suggested that the presence of Ti favors the superexchange (SE) interaction and suppresses the DE mechanism. Some authors attributed the decrease of $T_c$ to the variation of structural parameters such as the average Mn-O bond length $d_{\langle\mathrm{Mn-O}\rangle}$ and Mn-O-Mn bond angle $\theta_{\langle\mathrm{Mn-O-Mn}\rangle}$.\cite{Lakshmi,Kim} By calculating the $e_g$-electron bandwidth $W$ using an empirical formula, $W\propto cos[\frac{1}{2}(\pi-\theta_{\langle \mathrm{Mn-O-Mn}\rangle})]/d^{3.5}_{\langle\mathrm{Mn-O}\rangle}$, Kim \textit{et al.}\cite{Kim} observed a narrowing of $W$ in La$_{0.7}$Sr$_{0.3}$Mn$_{1-x}$Ti$_x$O$_3$ and related it to the decrease of $T_c$. Although $W(x)$ was found seemingly to vary in a consistent manner with $T_c(x)$ in $R_{1-x}A_x$MnO$_3$,\cite{Radaelli} explanations for the decrease of $T_c$ relying solely on the variation of $W(x)$ may not be relevant in Mn-site substituted manganites, especially when Mn is substituted by Al.\cite{Nam} To present the evidence for the inconsistent variation between $W(x)$ and $T_c(x)$, we plot in Fig. \ref{Fig4} the $W(x)$ curves calculated for the present La$_{0.7}$Ca$_{0.3}$Mn$_{1-x}$Al$_x$O$_3$ samples and also for the La$_{0.7}$Sr$_{0.3}$Mn$_{1-x}$Al$_x$O$_3$ samples used in Ref. 10. The results clearly show, for both sample series, a widening of the bandwidth with $x$, in contrast to the decrease of $T_c$ (see Fig. 5 and the results in Ref. 10). These results unambiguously indicate that the $e_g$-electron bandwidth alone does not account for the suppression of ferromagnetism and conductivity in magnetically diluted manganites.
\begin{figure}[t!]
  \includegraphics[width=6.3cm]{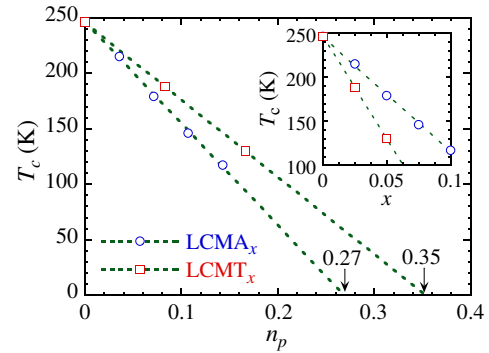}
  \caption{(Color online) The effect of selective dilution on $T_c$ of La$_{0.7}$Ca$_{0.3}$Mn$_{1-x}$\textit{M}$^\prime_x$O$_3$ at low dilution concentrations, $n_p$ ($n_p$=$x$/0.7 for LCMA$_x$ and $x$/0.3 for LCMT$_x$). The inset shows the original $T_c(x)$ data.}\label{Fig5}
\end{figure}

Nam \textit{et al.}\cite{Nam} have suggested that the variation of $T_c$ in diluted ferromagnetic manganites can be explained in terms of a molecular-field system with mixed interactions; $T_c$ can be written as
\begin{equation*}
T_\mathrm{c}=\frac{2S\left(S+1\right)}{3k_\mathrm{B}}\sum_{\alpha}^{n}z_{\alpha}J_{\alpha},
\label{eqn2}
\end{equation*}
where $k_\mathrm{B}$ is the Boltzmann's constant, $S$ the average
spin per magnetic ion, and $z_\alpha$ the number of ions involved in the interaction $\alpha$ with exchange coupling constant $J_\alpha$. If the system is slightly diluted by a non-magnetic element with concentration $n_p$, a linear dependence of $T_c$ is expected following the relation $z_\alpha$=$z_{\alpha0}$(1-$n_p$), where $z_{\alpha0}$ is the $z_{\alpha}$ value of the undiluted system. In mixed-valence manganites, while the Mn$^{3+}$-O$^{2-}$-Mn$^{4+}$ coupling (denoted as DE) is ferromagnetic according to the DE mechanism, the Mn$^{3+}$-O$^{2-}$-Mn$^{3+}$ (SE1) and Mn$^{4+}$-O$^{2-}$-Mn$^{4+}$ (SE2) couplings are superexchange antiferromagnetic. A selective substitution of Al$^{3+}$ (or Ti$^{4+}$) for Mn$^{3+}$ (or Mn$^{4+}$) causes both $z_\mathrm{DE}$ and $z_\mathrm{SE1}$ (or $z_\mathrm{SE2}$) to vary proportionally with 1-$n_p$ while $z_\mathrm{SE2}$ (or $z_\mathrm{SE1}$) is left intact. The previous study\cite{Nam} on La$_{0.7}$Sr$_{0.3}$Mn$_{1-x}$\textit{M}$^\prime_x$O$_3$ has found that, in the undiluted La$_{0.7}$Sr$_{0.3}$MnO$_3$ and slightly diluted compounds, both $J_\mathrm{SE2}$ and $J_\mathrm{SE1}$ are basically negligible, leading to a linear $T_c(n_p)$ dependence that crosses $T_c$=0 at $n_p$=1 scaled for both \textit{M}$^\prime$=Al and Ti. The absence of antiferromagnetic superexchange interactions is probably a reason behind the fact that La$_{0.7}$Sr$_{0.3}$MnO$_3$ has the highest $T_c$ ever found for manganites. If the antiferromagnetic superexchange interactions in the undiluted system are significant, $T_c$ will be reduced and the $T_c(n_p)$ linear dependence should intersect $T_c$=0 at a $n_p$ value smaller than 1. This seems to be qualitatively consistent with the results obtained for the present LCMA$_x$ and LCMT$_x$ systems presented in Fig. \ref{Fig5} where $T_c(n_p)$ curves are linear at low substitution concentrations and extrapolate to $T_c$=0 at $n_p$=0.27 for LCMA$_x$ and 0.35 for LCMT$_x$. The scaling behavior observed for La$_{0.7}$Sr$_{0.3}$Mn$_{1-x}$$M^\prime_x$O$_3$ is no longer relevant for La$_{0.7}$Ca$_{0.3}$Mn$_{1-x}$$M^\prime_x$O$_3$. Using the data in Fig. \ref{Fig5} and adopting $J_\mathrm{SE1}$=-0.58 meV derived from neutron scattering measurements on LaMnO$_3$,\cite{Moussa} we obtained $J_\mathrm{SE2}$=-4.34 meV and $J_\mathrm{DE}$=5.92 meV. These values may not be very accurate since $J_\mathrm{SE1}$ may change from LaMnO$_3$ to La$_{0.7}$Ca$_{0.3}$MnO$_3$ because of the difference in crystal structures, but essentially indicate the presence of significant antiferromagnetic interactions that coexist and compete with the DE FM one in La$_\mathrm{0.7}$Ca$_\mathrm{0.3}$MnO$_3$ and the slightly diluted compounds.
\section{Conclusion}
Selective dilution of Mn lattice in La$_{0.7}$Ca$_{0.3}$Mn$_{1-x}$\textit{M}$^\prime_x$O$_3$ (\textit{M}$^\prime$=Al,Ti) suppresses the ferromagnetism and metallic conduction, but strongly enhances the magnetoresistance. In contrast to the scaling behavior $T_c(n_p)$ previously reported for La$_{0.7}$Sr$_{0.3}$Mn$_{1-x}$\textit{M}$^\prime_x$O$_3$, extrapolations of the $T_c(n_p)$ linear curves to $T_c$=0 give different $n_p$ values and both much smaller than 1. These results indicate that, according to a molecular-field approximation, antiferromagnetic superexchange among Mn ions is significant in La$_{0.7}$Ca$_{0.3}$MnO$_3$. Our results suggest that variations of $e_g$-electron bandwidth $W$ alone cannot explain the decrease of $T_c$ in  Mn-site substituted manganites. We propose that this simple selective dilution technique can be quite effective in probing the presence of competing interactions in mixed systems such as manganites.
\begin{acknowledgments}
This work has been performed using facilities of the State Key Laboratories (IMS, VAST). The authors thank N. N. Toan and D. T. A. Thu for their help in redox titration experiments.
\end{acknowledgments}

\end{document}